\documentclass[english]{IOS-Book-Article}
\usepackage[T1]{fontenc}
\usepackage[latin9]{inputenc}
\usepackage{color}
\usepackage{amsthm}
\usepackage{amstext}
\usepackage{amssymb}
\usepackage{graphicx}

\makeatletter
  \theoremstyle{definition}
  \newtheorem{defn}{\protect\definitionname}
\theoremstyle{definition}
\newtheorem{assumption}{Assumption}

\usepackage{mathptmx}
\usepackage{cite}

\makeatother

\providecommand{\definitionname}{Definition}

\begin{document}
\begin{frontmatter}

\title{Navigation Function Based Decentralized Control of A Multi-Agent
System with Network Connectivity Constraints%
\thanks{Z. Kan and W. E. Dixon are with the Department of Mechanical and Aerospace
Engineering, University of Florida, Gainesville, FL, USA. Email: \{kanzhen0322,
wdixon\}@ufl.edu. John M. Shea is with the Department of Electrical
and Computer Engineering, University of Florida, Gainesville, USA.
Email: jshea@ece.ufl.edu. %
}%
\thanks{This research is supported in part by NSF award numbers 1161260, 1217908,
and a contract with the AFRL Mathematical Modeling and Optimization
Institute. Any opinions, findings and conclusions or recommendations
expressed in this material are those of the authors and do not necessarily
reflect the views of the sponsoring agency.%
}}

\maketitle
\begin{center}

\author{Zhen Kan, John M. Shea and Warren E. Dixon}

\end{center}
\begin{abstract}
A wide range of applications require or can benefit from collaborative
behavior of a group of agents. The technical challenge addressed in
this chapter is the development of a decentralized control strategy
that enables each agent to independently navigate to ensure agents
achieve a collective goal while maintaining network connectivity.
Specifically, cooperative controllers are developed for networked
agents with limited sensing and network connectivity constraints.
By modeling the interaction among the agents as a graph, several different
approaches to address the problems of preserving network connectivity
are presented, with the focus on a method that utilizes navigation
function frameworks. By modeling network connectivity constraints
as artificial obstacles in navigation functions, a decentralized control
strategy is presented in two particular applications, formation control
and rendezvous for a system of autonomous agents, which ensures global
convergence to the unique minimum of the potential field (i.e., desired
formation or desired destination) while preserving network connectivity.
Simulation results are provided to demonstrate the developed strategy.
\end{abstract}
\end{frontmatter}

\section{Introduction}

Multi-agent systems under cooperative control provide versatile platforms
for various commercial and military applications, such as formation
flight and cooperative attack in military systems \cite{Beard2006},
environmental sampling and distributed aperture observing for mobile
sensor networks \cite{zhang2010cooperative}, and intelligent highways
and air traffic control in transportation systems \cite{Tomlin1998a}.
These types of tasks usually require or can benefit from collaborative
motion of a group of agents, and thus the agents must be able to exchange
information over some form of communications network. For most applications,
communications will be over a wireless network, in which the communication
links between agents are dependent on the propagation of electromagnetic
signals between the agents, and the electromagnetic power density
decreases with distance. When performing desired tasks, the underlying
wireless communication can be impacted due to the motion of agents.
If the network is partitioned and the inter-agent communication is
disconnected, the agents can no longer coordinate their movements,
and the mission may fail. Hence, control algorithms must be designed
under the constraint of preserving network connectivity when performing
desired tasks.

\subsection{Overview of Research on Maintenance of Network Connectivity}

Network connectivity is a mainstream research focus. The interaction
of agents is typically modeled using constructs from graph theory,
and the graph determines which agents can exchange and share information
and how robust the group can behave in a dynamic environment. Proximity-based
graphs are generally used to capture the inter-agent communication.
In particular, a time varying graph $\mathcal{G}\left(t\right)$ is
used to model the dynamic graph, where $\mathcal{V}$ is the set of
vertices (representing the agents) and $\mathcal{E}(t)$ is the set
of edges connecting the vertices in $\mathcal{V}$. Each edge connecting
node $x$ and $y$ in $\mathcal{E}(t)$ specifies an available communication
link. 

A metric that is typically used to capture network connectivity is
the second smallest eigenvalue $\lambda_{2}\left(\mathcal{L}\right)$
of the Laplacian matrix $\mathcal{L}$ of the graph $\mathcal{G}$,
which is also known as the Fiedler value\cite{Godsil2001}. A positive
$\lambda_{2}\left(\mathcal{L}\right)$ indicates a connected graph,
and the associated eigenvector can be used to determine a set of links
that if removed will cause the network to partition \cite{RM_LAA:94}.
To ensure network connectivity, optimization based approaches are
developed in the works of \cite{kim2006} and \cite{DeGennaro2006}
to maximize the Fiedler value. However, the computation of $\lambda_{2}\left(\mathcal{L}\right)$
is generally centralized due to the requirement of the knowledge of
the entire network structure. Moreover, $\lambda_{2}\left(\mathcal{L}\right)$
is a non-differentiable function of the Laplacian matrix $\mathcal{L}$,
which presents an obstacle for designing continuous feedback controllers.
Alternative ways to overcome this constraint is to use the determinant
of $\mathcal{L}$ \cite{Zavlanos2007a}, which is a differentiable
function of $\mathcal{L}$, or achieving consensus on Laplacian eigenvectors
\cite{Yang2010}. 

Since the edge connection information is collected into the adjacency
matrix $A\left(\mathcal{G}\right)$, network connectivity can be captured
by the sum of powers of the adjacency matrix $\sum_{k=0}^{K}A^{k}$,
which represents the number of paths up to length $K$ connecting
two nodes in the graph $\mathcal{G}$ \cite{Godsil2001}. If every
entry in $\sum_{k=0}^{K}A^{k}$ is positive, any two nodes in the
graph $\mathcal{G}$ are connected with a path of maximum length $K$.
Following this idea, centralized optimization-based controllers are
developed in the work of \cite{Zavlanos2005} and \cite{srivastava2008}
to maintain the positiveness of all entries in $\sum_{k=0}^{K}A^{k}$.
Discrete-time approaches are discussed in \cite{Notarstefano2006,Bullo2009,Spanos2004,dimarogonas2008_R}
which rely on local gradients and switching of graphs in the case
of edge addition. This class of approaches are typically hybrid, since
both continuous edge preservation and discrete topology control are
considered.

Artificial potential fields based approaches that use attractive and
repulsive potentials are also widely utilized to guide the movement
of autonomous systems while preserving network connectivity. Particularly,
attractive potential fields are centered at the goal locations, and
repulsive potential fields are generated around obstacles. Driven
by the negative gradient of the potential field, each mobile robot
will converge to a minimum of the potential field, which is typically
the desired final position. Modeling network connectivity as an artificial
constraint, results such as \cite{Zavlanos2007a,Zavlanos2008,Zavlanos2009TAC,dimarogonas2008_R,ghaffarkhah_2009,dimarogonas2007,Olfati-Saber2004,Dimarogonas2010,meng2007}
are motivated by the need to prevent the graph from partitioning using
artificial potential fields. A potential field based centralized control
approach is developed in \cite{Zavlanos2007a} and \cite{Olfati-Saber2004}
to ensure the connectivity of a group of agents using the graph Laplacian
matrix. In \cite{Zavlanos2008}, connectivity control is performed
in the discrete space of graphs to verify link deletions with respect
to connectivity, and motion control is performed in the continuous
configuration space using a potential field. A potential field-based
neighbor control law is designed in \cite{Zavlanos2009TAC} to achieve
velocity alignment and network connectivity among different topologies.
In \cite{dimarogonas2008_R} and \cite{dimarogonas2007}, a repulsive
potential is used for a collision avoidance objective, and an attractive
potential field is used to drive agents together. Distributed control
laws are investigated to ensure edge maintenance in \cite{meng2007}
by allowing unbounded potential force whenever pairs of agents are
about to break existing links. In \cite{Dimarogonas2010}, a potential
field is designed for a group of mobile agents to perform desired
tasks while maintaining network connectivity; however, it is unclear
how the potential field method in \cite{Dimarogonas2010} can be extended
to include static obstacles. Other results that use artificial potential
fields for networked agents to perform formation control, rendezvous,
flocking and containment control while preserving network connectivity
include \cite{Kan2010a,Kan2010,Kan2011,Kan2011a,Kan.Klotz.ea2012,Kan.Pasiliao.ea2012,Kan.Shea.ea2012,Kan.Klotz.ea2013,Su2010a,Xiao.Wang.ea2012,Navaravong.Kan.ea2012}.

\subsection{Main Contributions}

A common problem with the aforementioned artificial potential field-based
control algorithms is the existence of local minima when attractive
and repulsive force are combined. When trapped by local minima, the
system will no longer converge to the desired minimum (i.e., control
objective) and result in mission failure. To avoid local minima, a
specific type of artificial potential, called a navigation function,
achieves a unique minimum (c.f., \cite{Rimon1992,Rimon_1990}) and
has been widely used in motion control for multi-agent systems. The
navigation function developed in \cite{Rimon_1990} is a real-valued
function that is designed so that the negated gradient field does
not have a local minima. The negated gradient of the navigation function
is attracted towards the goal and repulsed by obstacles for almost
all initial states. As such, closed-loop navigation function approaches
guarantee convergence to a desired destination. 

The development in the chapter utilizes ideas from navigation function
frameworks to control a group of agents with constraints on limited
sensing and network connectivity. Each agent is assumed to have limited
sensing capabilities or knowledge about the environment and limited
communication capabilities with nearby agents. To show the effectiveness
of the navigation function based approaches, two example applications,
formation control and rendezvous, developed in our previous works
of \cite{Kan.Dani.ea2012} and \cite{Kan2012submitted} are introduced.
In comparison to the above artificial potential field-based results,
the method developed in \cite{Kan.Dani.ea2012} achieves convergence
to a desired configuration and maintenance of network connectivity
using a decentralized navigation function approach which uses only
local feedback information. By using a local range sensor, an advantageous
feature of the developed decentralized controller is that no inter-agent
communication is required (i.e., communication free global decentralized
group behavior). That is, the goal is to maintain connectivity so
that radio communication is available when required for various task/mission
scenarios, but communication is not required to navigate, enabling
stealth modes of operation. In \cite{Kan2012submitted}, a group of
wheeled robots with nonholonomic constraints is tasked \textcolor{black}{with
the objective of rendezvousing at a common specified setpoint with
a desired orientation while maintaining network connectivity. Only
a subset of the robots are assumed to be aware of the global destination,
and the remaining robots must move with the constraint of ensuring
network connectivity so that the informed robots can guide the group
to the goal. Since the classical navigation function based approach
in }\cite{Kan2012submitted} is not applicable to robots with nonholonomic
constraints, \textcolor{black}{a decentralized time-varying continuous
controller is developed to reach the desired destination with a desired
orientation while preserving network connectivity based on a dipoloar
navigation function framework. Only local sensing feedback (i.e.,
relative position) from neighboring robots is used to navigate the
group. Simulation results demonstrate the performance of the developed
approaches.}

\section{Navigation Function Framework }

\begin{figure}
\centering{}\includegraphics[scale=0.5]{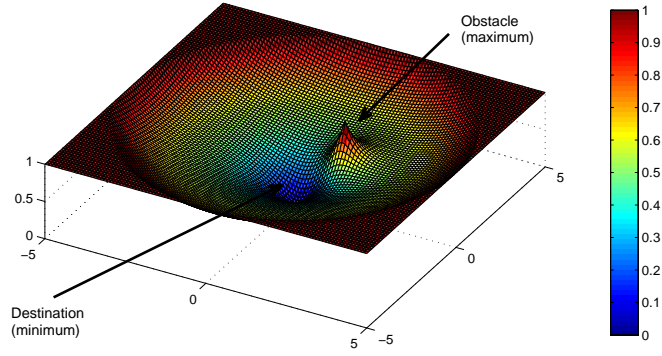}

\caption{An example of the artificial potential field generated for a disk-shaped
workspace with destination at the origin and an obstacle (i.e., artificial
constraints) located at $[1,1]^{T}$.}

\label{fig:potential}
\end{figure}

A navigation function is a particular category of potential functions
where the potential field does not have local minima and the negative
gradient vector field of the potential field guarantees almost global
convergence to a desired destination, along with (guaranteed) collision
avoidance, if the initial conditions do not lie within the sets of
measure zero. Formally, a navigation function is defined as:
\begin{defn}
\label{Def1}\cite{Rimon1992} \cite{Rimon_1990}Let $\mathcal{F}\subset E^{n}$
be a compact connected analytic manifold with a boundary. A map $\varphi:\mathcal{F\rightarrow}\left[0,1\right]$
is a Navigation Function, if it is: 1) smooth on $\mathcal{F}$ (at
least a $\mathcal{C}^{2}$ function); 2) admissible on $\mathcal{F}$,
(uniformly maximal on $\partial\mathcal{F}$ and constraint boundary);
3) polar on $\mathcal{F},$ ($q_{d}$ is a unique minimum); and 4)
a Morse function, (critical points of the navigation function are
non-degenerate).
\end{defn}
The second condition in Definition \ref{Def1} establishes that the
generated trajectories are collision-free, since the resulting vector
field is transverse to the boundary of $\mathcal{F}$, while the third
point indicates that, using a polar function on a compact connected
manifold with a boundary, all initial conditions are either brought
to a saddle point or to the unique minimum $q_{d}$. The requirement
that the navigation function is a Morse function ensures that the
initial conditions that bring the system to saddle points are sets
of measure zero \cite{Rimon_1990}. Given this property, all initial
conditions not within sets of measure zero are brought to the unique
minimum. An example of the generated artificial potential field is
shown in Fig. \ref{fig:potential} in which the destination is assigned
a minimum potential value, and the obstacle is assigned a maximum
potential value.

\section{Applications}

In this section, two results that are developed in our previous works
of \cite{Kan.Dani.ea2012} and \cite{Kan2012submitted} are discussed.
Based on the navigation function framework, a group of agents are
controlled to perform cooperative tasks, such as formation control
in \cite{Kan.Dani.ea2012} and rendezvous in \cite{Kan2012submitted},
while preserving network connectivity.

\subsection{Formation Control}

\subsubsection{Problem Formulation}

Consider a network composed of $N$ agents in the workspace $\mathcal{F}$,
where agent $i$ moves according to the following kinematics:
\begin{equation}
\dot{q}_{i}=u_{i},\text{ }i=1,\cdots,N\label{eq:ch3-dynamics}
\end{equation}
where $q_{i}\in\mathbb{R}^{2}$ denotes the position of agent $i$
in a two dimensional (2D) plane, and $u_{i}\in\mathbb{R}^{2}$ denotes
the velocity of agent $i$ (i.e., the control input). The workspace
$\mathcal{F}$ is assumed to be circular and bounded with radius $R,$
and $\partial\mathcal{F}$ denotes the boundary of $\mathcal{F}.$
Each agent in $\mathcal{F}$ is represented by a point-mass with a
limited communication and sensing capability encoded by a disk area.
Two moving agents can communicate with each other if they are within
a distance $R_{c}$, while the agent can sense stationary obstacles
or other agents within a distance $R_{s}$. For simplicity and without
loss of generality, assume that the sensing area coincides with the
communication area, i.e., $R_{c}=R_{s}$. A set of fixed points, $p_{1},\cdots,p_{M},$
are defined to represent $M$ stationary obstacles in the workspace
$\mathcal{F}$, and the index set of obstacles is denoted as $\mathcal{M=}\left\{ 1,\cdots,M\right\} $. 

The interaction of the system is modeled as a dynamic graph, in the
sense that the system evolves in time governed by the agent kinematics
in (\ref{eq:ch3-dynamics}). The dynamic graph is denoted as $\mathcal{G}(t)=(\mathcal{V},\mbox{ }\mathcal{E}(t))$,
where $\mathcal{V=}\left\{ 1,\cdots,N\right\} $ denotes the set of
nodes, and $\mathcal{E}(t)=\left\{ \left.\left(i,\mbox{ }j\right)\in\mathcal{V\times V}\right\vert d_{ij}\leq R_{c}\right\} $
denotes the set of time varying edges, where node $i$ and node $j$
are located at a position $q_{i}$ and $q_{j},$ and $d_{ij}\in\mathbb{R}^{+}$
is the relative distance defined as $d_{ij}=\left\Vert q_{i}-q_{j}\right\Vert $.
In graph $\mathcal{G}(t)$, each node $i$ represents an agent, and
the edge $\left(i,\mbox{ }j\right)$ denotes a link between agent
$i$ and $j$ when they stay within a distance $R_{c}$. The set of
neighbors of node $i$ (i.e., all the agents within the sensing zone
of agent $i$) is given by $\mathcal{N}_{i}=\left\{ \left.j,\text{\ }j\neq i\right\vert j\in\mathcal{V},\text{ }\left(i,\mbox{ }j\right)\in\mathcal{E}\right\} .$
One objective in this work is to have the multi-agent system converge
to a desired configuration, determined by a formation matrix $c_{ij}\in\mathbb{R}^{2}$
representing the desired relative position of node $i$ with an adjacent
node $j\in\mathcal{N}_{i}^{f}$, where $\mathcal{N}_{i}^{f}\subset\mathcal{N}_{i}$
denotes the set of nodes required to form a prespecified relative
position with node $i$ in the desired configuration. The neighborhood
$\mathcal{N}_{i}$ is a time varying set since nodes may enter or
leave the communication region of node $i$ at any time instant, while
$\mathcal{N}_{i}^{f}$ is a static set which is specified by the desired
configuration. The desired position of node $i$, denoted by $q_{di}$,
is defined as $q_{di}=\left\{ \left.q_{i}\right\vert \left\Vert q_{i}-q_{j}-c_{ij}\right\Vert ^{2}=0,\text{ }j\in\mathcal{N}_{i}^{f}\right\} .$
An edge $(i,\mbox{ }j)$ is only established between nodes $i$ and
$j$ if $j\in\mathcal{N}_{i}^{f}$.

A collision region \textit{}%
\footnote{\textit{\emph{The potential collision for node $i$ in this work not
only refers to the fixed obstacles, but also other moving nodes or
the workspace boundary, which are currently located in its collision
region.}}%
} is defined for each agent $i$ as a small disk with radius $\delta_{1}<R_{c}$
around the agent $i$, such that any other agent $j\in$ $\mathcal{N}_{i},$
or obstacle $p_{k},$ $k\in\mathcal{M}$, inside this region is considered
as a potential collision with agent $i$. To ensure connectivity,
an escape region for each agent $i$ is defined as the outer ring
of the communication area with radius $r,$ $R_{c}-\delta_{2}<r<R_{c},$
where $\delta_{2}\in\mathbb{R}$ is a predetermined buffer distance.
Edges formed with any node $j\in\mathcal{N}_{i}^{f}$ in the escape
region are in danger of breaking. The objective is to develop a decentralized
controller $u_{i}$ that uses relative position information from the
range sensor to regulate a connected initial graph to a desired configuration
while maintaining network connectivity and avoiding collisions with
other agents and obstacles.

\subsubsection{Control Design}

A decentralized controller is developed using only local sensing to
navigate the agents to a desired formation while maintaining network
connectivity. Consider a decentralized navigation function candidate
$\varphi_{i}:\mathcal{F}_{i}\rightarrow[0,1]$ for each node $i$
as
\begin{equation}
\varphi_{i}=\frac{\gamma_{i}}{\left(\gamma_{i}^{\alpha}+\beta_{i}\right)^{1/\alpha}},\label{eq: ch3-Navigation fcn}
\end{equation}
where $\alpha\in\mathbb{R}^{+}$ is a tuning parameter, $\gamma_{i}:\mathbb{R}^{2}\rightarrow\mathbb{R}^{+}$
is the goal function, and $\beta_{i}:\mathbb{R}^{2}\rightarrow[0,1]$
is a constraint function for node $i$. The goal function $\gamma_{i}$
in (\ref{eq: ch3-Navigation fcn}) encodes the control objective of
node $i$, specified in terms of the desired relative position with
respect to the adjacent nodes $\left\{ j\in\mathcal{N}_{i}^{f}\right\} $,
and drives the system to a desired configuration%
\footnote{The formation objective $\gamma_{i}$ is developed based on the desire
to control the distance and relative bearings between nodes. For some
applications, only the relative distance between nodes is important,
and the objective could be rewritten as $\gamma_{i}=\sum\nolimits _{j\in\mathcal{N}_{i}^{f}}\left(\left\Vert q_{i}-q_{j}\right\Vert -\left\Vert c_{ij}\right\Vert \right)^{2};$
however, this objective can introduce redundant desired configurations.
Future efforts could consider this alternative objective, where an
approach such as \cite{ghaffarkhah_2009} may be explored to address
the multiple desired minima.%
}. The goal function is designed as 
\begin{equation}
\gamma_{i}(q_{i},\text{ }q_{j})=\sum\nolimits _{j\in\mathcal{N}_{i}^{f}}\left\Vert q_{i}-q_{j}-c_{ij}\right\Vert ^{2}.\label{ch3-goal fcn}
\end{equation}
The constraint function $\beta_{i}$ in (\ref{eq: ch3-Navigation fcn})
is designed as 
\begin{equation}
\beta_{i}=B_{i0}\prod\nolimits _{j\in\mathcal{N}_{i}^{f}}b_{ij}\prod\nolimits _{k\in\mathcal{N}_{i}\cup\mathcal{M}_{i}}B_{ik},\label{eq:ch3-beta_i}
\end{equation}
to ensure collision avoidance and network connectivity by only accounting
for nodes and obstacles located within its sensing area during each
time instant. Specifically, the constraint function in (\ref{eq:ch3-beta_i})
is designed to vanish whenever node $i$ intersects with one of the
constraints in the environment, (i.e., if node $i$ touches a fixed
obstacle, the workspace boundary, other nodes, or departs away from
its adjacent nodes $\left\{ j\in\mathcal{N}_{i}^{f}\right\} $ to
a distance of $R_{c}$). 

In (\ref{eq:ch3-beta_i}), $b_{ij}\triangleq b(q_{i},\mbox{ }q_{j}):\mathbb{R}^{2}\rightarrow[0,1]$
ensures connectivity of the network graph (i.e., guarantees that nodes
$\left\{ j\in\mathcal{N}_{i}^{f}\right\} $ will never leave the communication
zone of node $i$ if node $j$ is initially connected to node $i$)
and is designed as\ 
\begin{equation}
b_{ij}=\left\{ \begin{array}{cc}
1 & d_{ij}\leq R_{c}-\delta_{2}\\
-\frac{1}{\delta_{2}^{2}}(d_{ij}+2\delta_{2}-R_{c})^{2}+\frac{2}{\delta_{2}}\left(d_{ij}+2\delta_{2}-R_{c}\right) & R_{c}-\delta_{2}<d_{ij}<R_{c}\\
0 & d_{ij}\geq R_{c}.
\end{array}\right.\label{eq:ch3-b_ij}
\end{equation}
Also in (\ref{eq:ch3-beta_i}), $B_{ik}\triangleq B(q_{i},\mbox{ }q_{k}):\mathbb{R}^{2}\rightarrow[0,1]$,
for point $k\in\mathcal{N}_{i}\cup\mathcal{M}_{i},$ where $\mathcal{M}_{i}$
indicates the set of obstacles within the sensing area of node $i$
at each time instant, ensures that node $i$ is repulsed from other
nodes or obstacles to prevent a collision, and is designed as
\begin{equation}
B_{ik}=\left\{ \begin{array}{cc}
-\frac{1}{\delta_{1}^{2}}d_{ik}^{2}+\frac{2}{\delta_{1}}d_{ik} & d_{ik}<\delta_{1}\\
1 & d_{ik}\geq\delta_{1}.
\end{array}\right.\label{eq:ch3-B_ik}
\end{equation}
Similarly, the function $B_{i0}$ in (\ref{eq:ch3-beta_i}) is used
to model the potential collision of node $i$ with the workspace boundary,
where the positive scalar $B_{i0}\in\mathbb{R}$ is designed similar
to $B_{ik}$ by replacing $d_{ik}$ with $d_{i0}$, where $d_{i0}\in\mathbb{R}^{+}$
is the relative distance of the node $i$ to the workspace boundary
defined as $d_{i0}=R-\left\Vert q_{i}\right\Vert $.

Based on the definition of the navigation function candidate, the
decentralized controller for each node is designed as
\begin{equation}
u_{i}=-K\nabla_{q_{i}}\varphi_{i},\label{ch3-controller}
\end{equation}
where $K$ is a positive gain, and $\nabla_{q_{i}}\varphi_{i}$ is
the gradient of $\varphi_{i}$ with respect to $q_{i}$. Hence, the
controller in (\ref{ch3-controller}) is bounded and yields the desired
performance by steering node $i$ along the direction of the negative
gradient of $\varphi_{i}$ if (\ref{eq: ch3-Navigation fcn}) is a
navigation function. Due to space limitation, the proof that (\ref{eq: ch3-Navigation fcn})
is a qualified navigation function is not included and can be referred
to the work of \cite{Kan.Dani.ea2012}.

\subsubsection{Simulation Results}

\begin{figure}
\centering{}\includegraphics[scale=0.4]{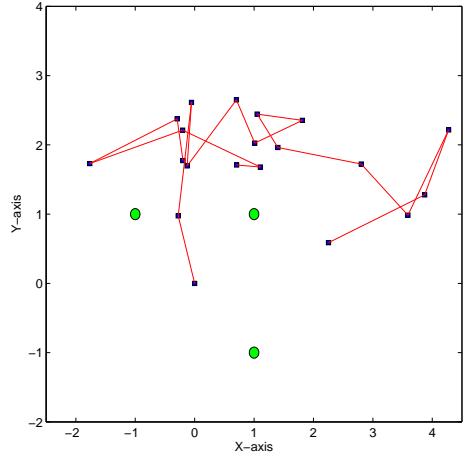}

\caption{A connected initial graph with desired neighborhood in the workspace
with static obstacles, where dots represent the static obstacles,
squares represent agents, and the line connecting the nodes indicate
the available communication between nodes.}

\label{fig:uf_initial}
\end{figure}

\begin{figure}
\centering{}\includegraphics[scale=0.4]{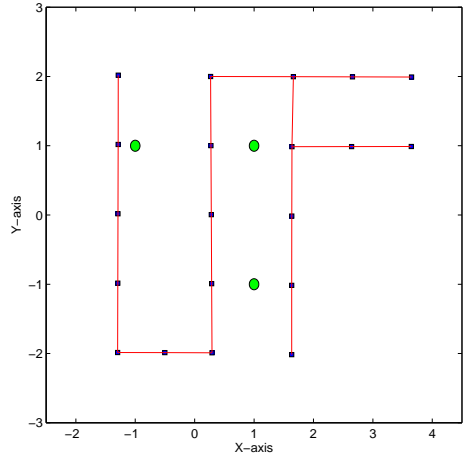}

\caption{The achieved final configuration.}

\label{fig:uf_final}
\end{figure}

\begin{figure}
\centering{}\includegraphics[scale=0.4]{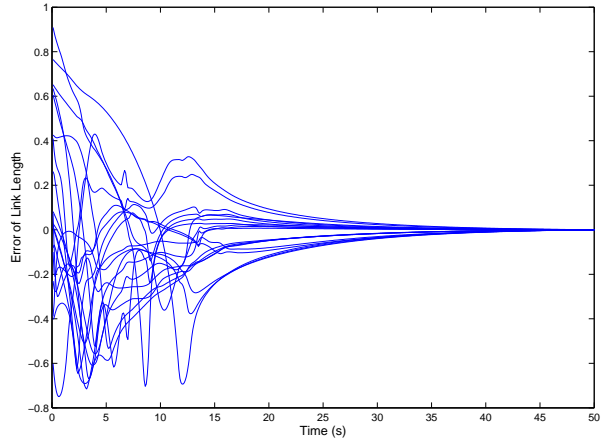}

\caption{The error plot.}

\label{fig:uf_error}
\end{figure}

\begin{figure}
\centering{}\includegraphics[scale=0.4]{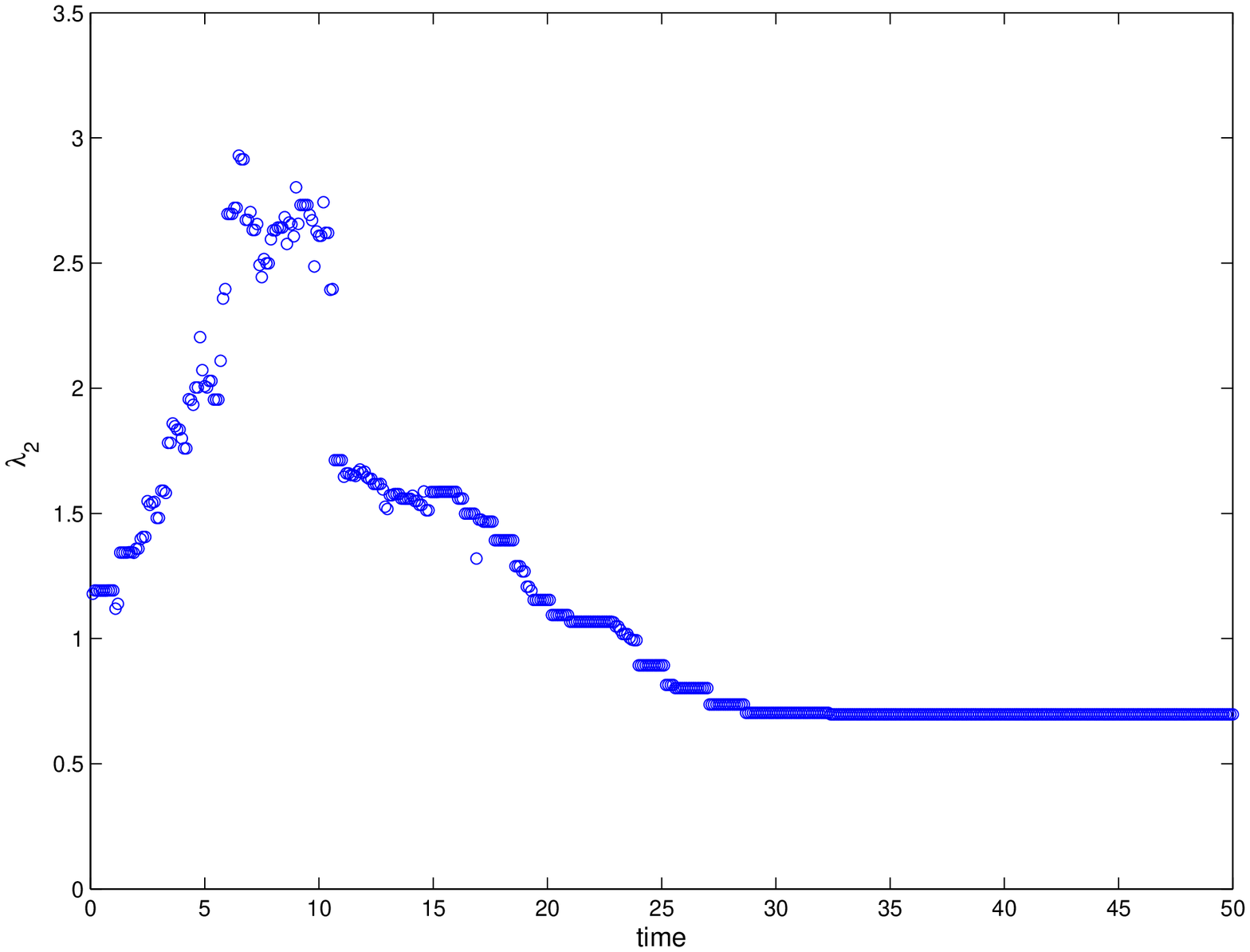}

\caption{The plot of the Fiedler eigenvalue of the Laplacian matrix during
the evolution. The circle indicates the Fiedler eigenvalue of the
graph at each time instance.}

\label{fig:uf_lamda}
\end{figure}

Simulation results illustrate the performance of the proposed control
strategy. As shown in the Fig. \ref{fig:uf_initial}, a connected
initial graph of 20 nodes with kinematics in (\ref{eq:ch3-dynamics})
are randomly deployed with desired neighborhood in a workspace of
$R=10$ m with static obstacles. Each node is assumed to have a limited
communication and sensing zone of $R_{c}=2$ m. The squares and dots
denote the moving agents and the static obstacles respectively, while
the solid line connecting two nodes represents a communication link,
indicating that the two agents are located within each other's communication
and sensing zone. The desired configuration is characterized by a
shape of {}``UF''. The system is simulated for $50s$ with the step
size of $0.1$. The tuning parameter $\alpha$ in (\ref{eq: ch3-Navigation fcn})
is set as $\alpha=1.5$, and $\delta_{1}=\delta_{2}=0.4$ m in (\ref{eq:ch3-b_ij})
and (\ref{eq:ch3-B_ik}). Results in Fig. \ref{fig:uf_final} indicate
that the system finally converges to the desired configuration. Fig.
\ref{fig:uf_error} shows the inter-node distance between nodes converges
to the desired value. To show the connectivity of the network during
the evolution, the Fiedler eigenvalue of the graph Laplacian matrix
is plotted in Fig. \ref{fig:uf_lamda}. Since the Fiedler eigenvalue
is always positive, the graph is connected \cite{Godsil2001}.


\subsection{Rendezvous for Mobile Agents with Nonholonomic Constraints}

\subsubsection{Problem Formulation}

Consider $N$ networked mobile robots operating in a workspace $\mathcal{F}$,
where $\mathcal{F}$ is a bounded disk area with radius $R_{w}$.
Each robot in $\mathcal{F}$ moves according to the following nonholonomic
kinematics:
\begin{equation}
\dot{q}_{i}=\left[\begin{array}{cc}
\cos\theta_{i} & 0\\
\sin\theta_{i} & 0\\
0 & 1
\end{array}\right]\left[\begin{array}{c}
v_{i}\left(t\right)\\
\omega_{i}\left(t\right)
\end{array}\right],\text{ }i=1,\cdots,N\label{eq:dynamics}
\end{equation}
where $q_{i}\left(t\right)\triangleq\left[\begin{array}{cc}
p_{i}^{T}\left(t\right) & \theta_{i}\left(t\right)\end{array}\right]^{T}\in\mathbb{R}^{3}$ denotes the states of robot $i,$ with $p_{i}\left(t\right)\triangleq\left[\begin{array}{cc}
x_{i}\left(t\right) & y_{i}\left(t\right)\end{array}\right]^{T}\in\mathbb{R}^{2}$ denoting the position of robot $i$, and $\theta_{i}\left(t\right)\in\left(-\pi,\pi\right]$
denoting the robot orientation with respect to the global coordinate
frame in $\mathcal{F}$. In (\ref{eq:dynamics}), $v_{i}\left(t\right),$
$\omega_{i}\left(t\right)\in\mathbb{R}$ are the control inputs that
represent the linear and angular velocity of robot $i,$ respectively.

\textcolor{black}{Assume that each robot has sensing and communication
limitations encoded by a disk area with radius $R,$ which indicates
that two moving robots can sense and communicate with each other as
long as they stay within a distance of $R.$ We also assume that only
a subset of the robots, called informed robots, are provided with
knowledge of the destination, while the other robots can only use
local state feedback (i.e., position feedback from immediate neighbors
and absolute orientation measurement). Furthermore, while multiple
informed robots may be used for rendezvous, the analysis and results
of this work are focused on a single informed robot. The techniques
proposed in this work could be extended to the case of multiple informed
robots by using containment control \cite{Kan.Klotz.ea2013,Kan.Mehta.ea2014,Cao2009}.
}The interaction among the robots is modeled as a directed graph $\mathcal{G}\left(t\right)=\left(\mathcal{V},\mathcal{E}(t)\right)$,
where the node set $\mathcal{V=}\left\{ 1,\cdots,N\right\} $ represents
the group of robots, and the edge set $\mathcal{E}(t)$ denotes time-varying
edges. The set of informed robots and followers are denoted as $\mathcal{V}_{L}$
and $\mathcal{V}_{F}$, respectively, such that $\mathcal{V}_{L}\cup\mathcal{V}_{F}=\mathcal{V}$
and $\mathcal{V}_{L}\cap\mathcal{V}_{F}=\emptyset$. Let $\mathcal{V}_{L}=\left\{ 1\right\} $
and $\mathcal{V}_{F}=\left\{ 2,\cdots,N\right\} $. A directed edge
$\left(i,j\right)\in$ $\mathcal{E}$ in $\mathcal{G}\left(t\right)$
exists between node $i$ and $j$ if their relative distance $d_{ij}\triangleq\left\Vert p_{i}-p_{j}\right\Vert \in\mathbb{R}^{+}$
is less than $R$. The directed edge $\left(i,j\right)$ indicates
that node $i$ is able to access the states (i.e., position and orientation)
of node $j$ through local sensing, but not vice versa$.$ Accordingly,
node $j$ is a neighbor of node $i$ (also called the parent of node
$i$), and the neighbor set of node $i$ is denoted as $\mathcal{N}_{i}=\left\{ j\text{ }|\text{ }\left(i,j\right)\in\mathcal{E}\right\} $,
which includes the nodes that can be sensed. A directed spanning tree
is a directed graph, where every node has one parent except for one
node, called the root, and the root node has directed paths to every
other node in the graph. Since the follower robots are not aware of
the destination, they have to stay connected with the informed robot
either directly or indirectly through concatenated paths, such that
the knowledge of the destination can be delivered to all the nodes
through the connected network. Hence, to complete the desired tasks,
maintaining connectivity of the underlying graph is necessary. 

The main objectives are to derive a set of distributed controllers
using only local information (i.e., the position feedback from other
robots within a sensing area) to lead the robots to rendezvous at
a common destination $p^{\ast}$ with a desired orientation $\theta^{\ast},$
i.e., $q_{i}^{\ast}=\left[\begin{array}{cc}
\left(p^{\ast}\right)^{T} & \theta^{\ast}\end{array}\right]^{T}$ $\forall i$ in the workspace $\mathcal{F},$ while guaranteeing
the underlying graph $\mathcal{G}\left(t\right)$ remains connected
during the system evolution, provided the given initial graph has
a directed spanning tree. 
\begin{assumption}
\label{Ass1}The initial graph $\mathcal{G}\left(0\right)$ has a
directed spanning tree with the informed node as the root.
\end{assumption}

\subsubsection{Control Design}

\begin{figure}
\centering{}\includegraphics[scale=0.4]{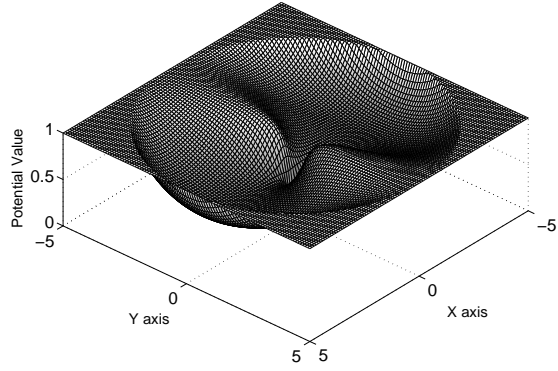}

\caption{An example of a dipolar navigation function with a workspace of $R_{w}=5$
and destination located at the origin with a desired orientation $\theta^{\ast}=0$.}

\label{fig:dipolar}
\end{figure}

In contrast to the fully actuated dynamics in (\ref{eq:ch3-dynamics}),
mobile agents with nonholonomic constraints in (\ref{eq:dynamics})
are considered. The navigation function introduced in \cite{Rimon1992}
and \cite{Rimon_1990} ensures global convergence of the closed-loop
system; however, the approach is not suitable for nonholonomic systems,
since the feedback law generated from the gradient of the navigation
function can lead to undesirable behavior. To overcome the undesirable
behaviors, the original navigation function was extended to a Dipolar
Navigation Function in \cite{Tanner2000} and \cite{Tanner2003},
where the flow lines created in the potential field resemble a dipole,
so that the flow lines are all tangent to the desired orientation
at the origin and the vehicle can achieve the desired orientation.
One example of the dipolar navigation is shown in Fig. \ref{fig:dipolar},
where the potential field has a unique minimum at the destination
(i.e., $p^{\ast}=\left[0,0\right]^{T}$ and $\theta^{\ast}=0$)$,$
and achieves the maximums at the workspace boundary of $R_{w}=5$.
Note that the surface $x=0$ divides the workspace into two parts
and forces all the flow lines to approach the destination parallel
to the $y$-axis.

The control strategy is to develop a dipolar navigation function for
the informed robot, which creates a feasible nonholonomic trajectory
for the nonholonomic robot and guarantees the achievement of the specified
destination with a desired orientation, while other follower robots
aim to achieve consensus with the informed robot and maintain network
connectivity by using only local interaction with neighboring robots.
Following this idea, the dipolar navigation function is designed for
the informed node $i\in\mathcal{V}_{L}$ as $\varphi_{i}^{d}\left(t\right):\mathcal{F}\rightarrow[0,1],$
\begin{equation}
\varphi_{i}^{d}=\frac{\gamma_{d}}{\left(\gamma_{d}^{\alpha}+H_{d}\cdot\beta_{d}\right)^{1/\alpha}},\label{phi_d}
\end{equation}
where $\alpha\in\mathbb{R}^{+}$ is a tuning parameter. The goal function
$\gamma_{d}\left(t\right):\mathbb{R}^{2}\rightarrow\mathbb{R}^{+}$
in (\ref{phi_d}) encodes the control objective of achieving the desired
destination, specified by the distance from $p_{i}\left(t\right)\in\mathbb{R}^{2}$
to the destination $p^{\ast}\in\mathbb{R}^{2},$ which is designed
as 
\[
\gamma_{d}=\left\Vert p_{i}\left(t\right)-p^{\ast}\right\Vert ^{2}.
\]
The factor $H_{d}\left(t\right)\in\mathbb{R}^{+}$ in (\ref{phi_d})
creates a repulsive potential to align the trajectory of node $i$
at the destination with the desired orientation. The repulsive potential
factor is designed as
\begin{equation}
H_{d}=\varepsilon_{nh}+\left(\left(p_{i}-p^{\ast}\right)^{T}\cdot n_{d}\right)^{2},\label{dipolar}
\end{equation}
where $\varepsilon_{nh}$ is a small positive constant, and $n_{d}=\left[\begin{array}{cc}
\cos\left(\theta^{\ast}\right) & \sin\left(\theta^{\ast}\right)\end{array}\right]^{T}\in\mathbb{R}^{2}.$ A small disk area with radius $\delta_{1}<R$ centered at node $i$
is denoted as a collision region. To prevent a potential collision
between node $i$ and the workspace boundary $\partial\mathcal{F}$,
the function $\beta_{d}:\mathbb{R}^{2}\rightarrow\left[0,1\right]$
in (\ref{phi_d}) is designed as
\begin{equation}
\beta_{d}=\left\{ \begin{array}{cc}
-\frac{1}{\delta_{1}^{2}}d_{i0}^{2}+\frac{2}{\delta_{1}}d_{i0}, & d_{i0}<\delta_{1}\\
1, & d_{i0}\geq\delta_{1},
\end{array}\right.\label{eq:B_i0}
\end{equation}
where $d_{i0}\triangleq R_{w}-\left\Vert p_{i}\right\Vert \in\mathbb{R}$
is the relative distance of node $i$ to the workspace boundary.

Since $\gamma_{d}$ and $\beta_{d}$ in (\ref{phi_d}) are guaranteed
to not be zero simultaneously, the navigation function candidate in
(\ref{phi_d}) achieves its minimum of $0$ when $\gamma_{d}=0$ and
achieves its maximum of $1$ when $\beta_{d}=0$. Our previous work
in \cite{Kan.Dani.ea2012} proves that the original navigation function
with the form of $\varphi_{i}=\frac{\gamma_{i}}{\left(\gamma_{i}^{\alpha}+\beta_{i}\right)^{1/\alpha}}$
is a qualified navigation function. It is also shown in \cite{loizou2008}
that the navigation properties are not affected by the modification
to a dipolar navigation with the design of (\ref{dipolar}), as long
as the workspace is bounded, $H_{d}$ in (\ref{phi_d}) can be bounded
in the workspace, and $\varepsilon_{nh}$ is a small positive constant.
As a result, the decentralized navigation function $\varphi_{i}^{d}$
proposed in (\ref{phi_d}) can be proven to be a qualified navigation
function by following a similar procedure in \cite{loizou2008} and
\cite{Kan.Dani.ea2012}. From the properties of the navigation function,
it is known that almost all initial positions (except for a set of
measure zero points) asymptotically approach the desired destination.

To track the informed node while maintaining network connectivity,
a local interaction rule is designed for each follower node $i\in\mathcal{V}_{F}$
as $\varphi_{i}^{f}\left(t\right):\mathcal{F}\rightarrow[0,1],$
\begin{equation}
\varphi_{i}^{f}=\frac{\gamma_{i}}{\left(\gamma_{i}^{\alpha}+\beta_{i}\right)^{1/\alpha}},\label{phi_f}
\end{equation}
where $\alpha\in\mathbb{R}^{+}$ is a tuning parameter. The goal function
$\gamma_{i}\left(t\right):\mathbb{R}^{2}\rightarrow\mathbb{R}^{+}$
in (\ref{phi_f}) encodes the control objective of achieving consensus
on the position between node $i$ and neighboring nodes $j\in\mathcal{N}_{i}$,
which is designed as 
\begin{equation}
\gamma_{i}=\sum_{j\in\mathcal{N}_{i}}\left\Vert p_{i}\left(t\right)-p_{j}\left(t\right)\right\Vert ^{2}.\label{goal fcn}
\end{equation}
To ensure connectivity of the existing links between nodes $i$ and
its neighboring nodes $j\in\mathcal{N}_{i}$, an escape region for
each node is defined as the outer ring of the sensing area with radius
$r,$ $R-\delta_{2}<r<R,$ where $\delta_{2}\in\mathbb{R}^{+}$ is
a predetermined buffer distance. Each edge formed by node $i$ and
the adjacent node $j\in\mathcal{N}_{i}$ in the escape region have
the potential to break connectivity. Hence, the constraint function
$\beta_{i}:\mathbb{R}^{2N}\rightarrow\left[0,1\right]$ in (\ref{phi_f})
is designed as 
\begin{equation}
\beta_{i}=\prod\nolimits _{j\in\mathcal{N}_{i}}b_{ij},\label{beta}
\end{equation}
where $b_{ij}\triangleq b(p_{i},$ $p_{j}):\mathbb{R}^{2}$ $\rightarrow\left[0,1\right]$
ensures connectivity of the existing links between nodes $i$ and
its neighboring nodes $j\in\mathcal{N}_{i}$ (i.e., guarantees that
nodes $j\in\mathcal{N}_{i}$ will never leave the sensing and communication
zone of node $i$ if node $j$ is initially connected to node $i$)
and is designed as\ 
\begin{equation}
b_{ij}=\left\{ \begin{array}{cc}
1, & d_{ij}\leq R-\delta_{2}\\
\begin{array}{c}
\begin{array}{c}
-\frac{1}{\delta_{2}^{2}}(d_{ij}+2\delta_{2}-R)^{2}\\
+\frac{2}{\delta_{2}}\left(d_{ij}+2\delta_{2}-R\right),
\end{array}\end{array} & R-\delta_{2}<d_{ij}<R\\
0, & d_{ij}\geq R.
\end{array}\right.\label{eq:b_ij}
\end{equation}
The constraint function in (\ref{beta}) is designed to vanish whenever
node $i$ meets the constraints of network connectivity in the workspace,
(i.e., if node $i$ departs from its neighbor nodes $j\in\mathcal{N}_{i}$
to a distance of $R$). Since $\gamma_{i}$ and $\beta_{i}$ in (\ref{phi_f})
will not be zero simultaneously from their definitions, it is clear
that $\varphi_{i}^{f}$ achieves its minimum of $0$ if $\gamma_{i}=0$
(i.e., the consensus is reached between node $i$ and its immediate
neighbors), and $\varphi_{i}^{f}$ achieves its maximum of $1$ if
$\beta_{i}=0$ (i.e., the constraint of network connectivity is met).

For brevity, $\varphi_{i}$ is used to represent the potential function
designed for each node $i$, where particularly $\varphi_{i}=\varphi_{i}^{d}$
in (\ref{phi_d}) if $i\in\mathcal{V}_{L}$, and $\varphi_{i}=\varphi_{i}^{f}$
in (\ref{phi_f}) if $i\in\mathcal{V}_{F}.$ The desired orientation
for any robot $i\in\mathcal{V}$, denoted by $\theta_{di}\left(t\right),$
is defined as a function of the negative gradient of the decentralized
function $\varphi_{i}$ as,
\begin{equation}
\theta_{di}\triangleq\arctan2\left(\begin{array}{cc}
-\frac{\partial\varphi_{i}}{\partial y_{i}}, & -\frac{\partial\varphi_{i}}{\partial x_{i}}\end{array}\right),\label{theta_d}
\end{equation}
where $\arctan2\left(\cdot\right):\mathbb{R}^{2}\rightarrow\mathbb{R}$
denotes the four quadrant inverse tangent function, and $\theta_{di}\left(t\right)$
is confined to the region of $\left(-\pi,\pi\right]$. By defining
$\theta_{di}\left\vert _{p^{\ast}}\right.=\arctan2\left(0,0\right)=\theta_{i}\left\vert _{p^{\ast}}\right.$,
$\theta_{di}$ remains continuous along any approaching direction
to the goal position. Based on the definition of $\theta_{di}$ in
(\ref{theta_d})
\begin{equation}
\nabla_{i}\varphi_{i}=-\left\Vert \nabla_{i}\varphi_{i}\right\Vert \left[\begin{array}{cc}
\cos\left(\theta_{di}\right) & \sin\left(\theta_{di}\right)\end{array}\right]^{T},\label{gradient}
\end{equation}
where $\nabla_{i}\varphi_{i}=\left[\begin{array}{cc}
\frac{\partial\varphi_{i}}{\partial x_{i}} & \frac{\partial\varphi_{i}}{\partial y_{i}}\end{array}\right]^{T}$ denotes the partial derivative of $\varphi_{i}$ with respect to
$p_{i}$, and $\left\Vert \nabla_{i}\varphi_{i}\right\Vert $ denotes
the Euclidean norm of $\nabla_{i}\varphi_{i}$. The difference between
the current orientation and the desired orientation for robot $i$
at each time instant is defined as 
\begin{equation}
\tilde{\theta}_{i}\left(t\right)=\theta_{i}\left(t\right)-\theta_{di}\left(t\right),\label{error}
\end{equation}
where $\theta_{di}\left(t\right)$ is generated from the decentralized
navigation function $\varphi_{i}$ and (\ref{theta_d}). 

Based on the open-loop system in (\ref{eq:dynamics}), the controller
for each robot (i.e., the linear and angular velocity of robot $i$)
is designed as
\begin{equation}
v_{i}=k_{v,i}\left\Vert \nabla_{i}\varphi_{i}\right\Vert \cos\tilde{\theta}_{i},\label{v_linear}
\end{equation}
\begin{equation}
\omega_{i}=-k_{w,i}\tilde{\theta}_{i}+\dot{\theta}_{di},\label{v_angular}
\end{equation}
where $k_{v,i}$, $k_{w,i}$$\in\mathbb{R}^{+}$ denote the control
gains for robot $i$.

\subsubsection{Simulation Results}

\begin{figure}
\centering{}\includegraphics[scale=0.4]{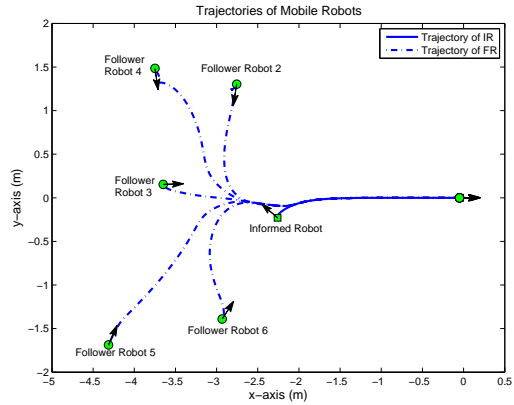}

\caption{Plot of robot trajectories with solid line and dot-dash line indicating
the trajectory of the informed robot (IR) and the follower robot (FR),
respectively.}

\label{fig:trajectory}
\end{figure}

\begin{figure}
\centering{}\includegraphics[scale=0.4]{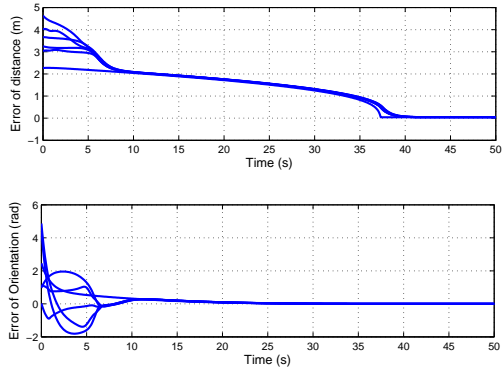}

\caption{Error plot of the distance to the destination and the error plot of
the orientation $\theta_{i}-\theta^{\ast}$.}

\label{fig:error}
\end{figure}

\begin{figure}
\centering{}\includegraphics[scale=0.4]{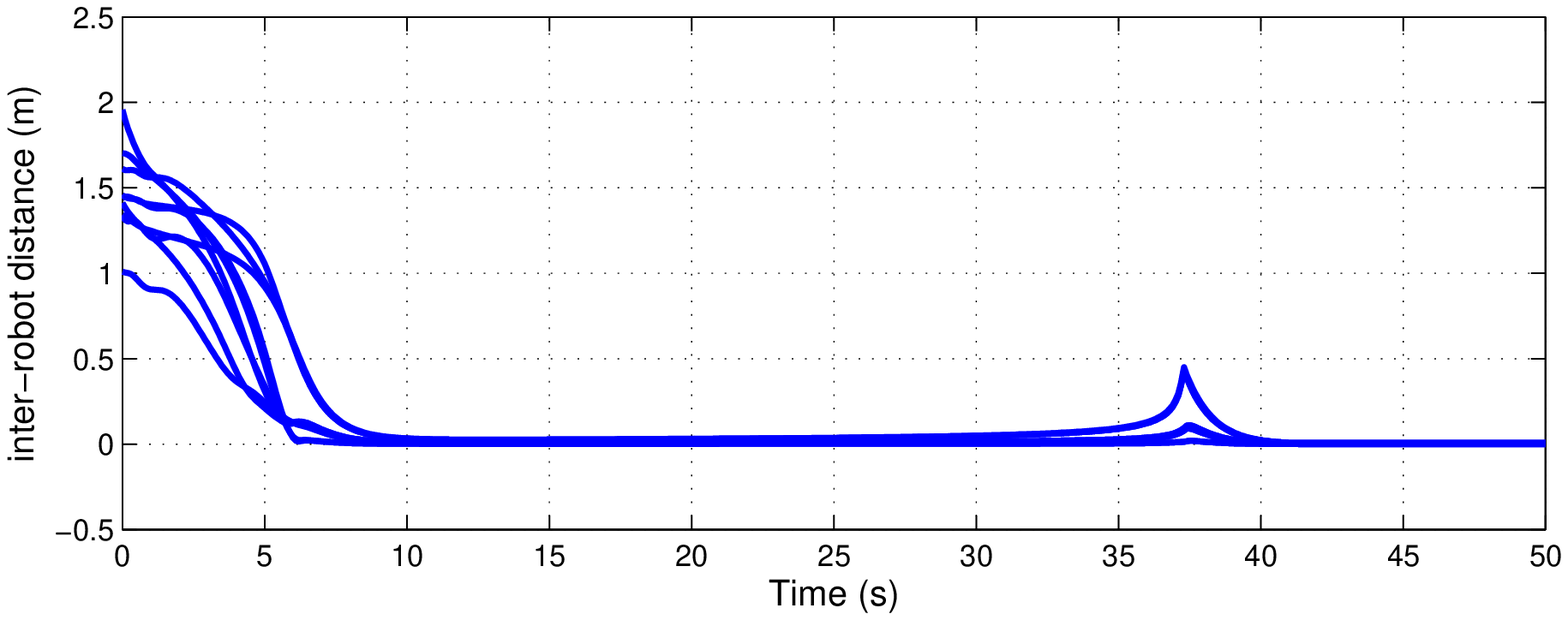}

\caption{The evolution of inter-robot distance.}

\label{fig:dis}
\end{figure}

The following numerical simulation demonstrates the performance of
the controller developed in (\ref{v_linear}) and (\ref{v_angular})
in a scenario in which a group of six mobile robots with the kinematics
in (\ref{eq:dynamics}) are navigated to the common destination $p^{\ast}=\left[\begin{array}{cc}
0 & 0\end{array}\right]^{T}$ with the desired orientation $\theta^{\ast}=0.$ The limited communication
and sensing zone for each robot is assumed as $R=2$ m and $\delta_{1}=\delta_{2}=0.4$
m. The tuning parameter $\alpha$ in (\ref{phi_d}) is selected as
$\alpha=1.2$. The group of mobile robots is arbitrarily deployed
in the workspace and forms a connected graph. The informed node is
randomly selected from the group, and is the only node aware of the
desired destination $p^{\ast}$ and orientation $\theta^{\ast}$.
The control laws in (\ref{v_linear}) and (\ref{v_angular}) yield
the simulation results shown in Fig. \ref{fig:trajectory}-\ref{fig:dis}.
Fig. \ref{fig:trajectory} shows the trajectory for each robot, where
the associated arrows indicate the initial or final orientation. In
Fig. \ref{fig:error}, the position and orientation error plot indicates
that each robot achieves the common destination with the desired orientation.
The evolution of the inter-robot distance is shown in Fig. \ref{fig:dis},
which implies that the connectivity of the underlying graph is maintained,
since the inter-robot distance is less than the radius $R=2$ m during
the motion.


\section{Conclusion and Future Work}

An overview of the current research in preserving network connectivity
for networked agents is provided, with a focus on the use of navigation
function framework in two particular applications, formation control
and rendezvous, for agents with limited sensing and network connectivity
constraints. Prior works based on artificial potential fields can
cause the system to be trapped by local minima, and thus result in
mission failure. By modeling the network connectivity as an artificial
obstacle in navigation functions, the developed control strategy ensures
global convergence to the unique minimum of the potential field (i.e.,
control objective) while maintaining network connectivity. 

In the formation control result, the initial topology is assumed to
be a supergraph of the desired topology, which ensures that the agents
are originally in a feasible interconnected state. Additional efforts
could consider formation control from an arbitrary initial graph to
a desired graph. Additional efforts could also incorporate more realism
into the physical and communications models, by accounting for the
dynamics of the robots and the effects of those on communications,
and incorporating more realistic channel models. In the rendezvous
result, although robots are guaranteed to converge to the desired
destination, the rate of convergence is not considered. Generally,
the rate of convergence depends on the network topology, which is
a function of the roles of nodes (i.e., informed nodes or followers)
and their interactions. \textcolor{black}{A different set of informed
nodes may lead to different convergence rates. Extension of this work
could seek to optimize performance metrics such as the degree of connectivity
and the convergence rate of the network in scenarios where the set
of informed nodes can be determined and/or positioned a priori.}

\bibliographystyle{IEEEtran}
\bibliography{master,ncr,\string"E:/nonliner control/my paper/bibtex/bib/ncrbibs/master\string",\string"E:/nonliner control/my paper/bibtex/bib/ncrbibs/ncr\string",\string"C:/ZHEN KAN/NCR/My Paper/bibtex/bib/ncrbibs/master\string",\string"C:/ZHEN KAN/NCR/My Paper/bibtex/bib/ncrbibs/ncr\string"}

\end{document}